\documentclass[journal]{IEEEtran}
\usepackage{amsmath}
\usepackage{amssymb}
\usepackage{amsmath,amssymb,amsfonts}
\usepackage{graphicx}
\usepackage{booktabs}
\usepackage{multirow}
\usepackage{array}
\usepackage{xcolor}
\usepackage{cite}
\usepackage{url}
\usepackage{placeins}
\usepackage{algorithm}
\usepackage{algpseudocode}
\usepackage{tikz}
\usetikzlibrary{positioning,arrows.meta,fit,calc}
\usepackage{stfloats}

\begin{document}

\title{PeakFlow: Peak-Guided Coarse-to-Refined Modeling for EEG-Based Dynamic Affective Trajectory Prediction}

\author{Hao Tang, Songyun Xie\textsuperscript{*}, \IEEEmembership{Member, IEEE}, Xinzhou Xie\textsuperscript{*}, Can Liao, Xin Zhang, Bohan Li, 

Zhongyu Tian, 
Dalu Zheng

\thanks{This work was supported by the National Natural Science Foundation of China (No. 62220106007). (Co-corresponding authors: Songyun Xie, Xinzhou Xie)}
\thanks{Hao Tang, Can Liao, Zhongyu Tian, and Dalu Zheng are with the School of Electronics and Information,
Northwestern Polytechnical University, Xi'an, Shaanxi 710072, China (e-mail: haotang@mail.nwpu.edu.cn; liao\_can@mail.nwpu.edu.cn; tzy2157@mail.nwpu.edu.cn; zhengdl@mail.nwpu.edu.cn).

Songyun Xie, Xinzhou Xie, and Xin Zhang are with the School of Artificial Intelligence,
Northwestern Polytechnical University, Xi'an, Shaanxi 710072, China (e-mail: syxie@nwpu.edu.cn; xinzhxie@nwpu.edu.cn; xzhang@nwpu.edu.cn).

Bohan Li is with the Institute of Medical Research,
Northwestern Polytechnical University, Xi'an, Shaanxi 710072, China (e-mail: bhli@mail.nwpu.edu.cn).

}

}

\maketitle

\begin{abstract}
Most existing EEG-based emotion recognition studies formulate affective decoding as static category prediction.
However, emotions elicited by continuous stimulation evolve over time, gradually accumulate, reach peak intensity, and then recover.
This motivates EEG-based dynamic affective trajectory prediction, where the goal is to estimate continuous affective intensity curves from sequential EEG observations.
Existing temporal regression models can capture coarse intensity trends, but they often fail to preserve peak-centered temporal structure, leading to inaccurate peak timing and terminal-peak bias, where the predicted maximum intensity is incorrectly shifted toward the end of a trial.
To address this issue, we propose \emph{PeakFlow}, a peak-guided coarse-to-refined framework for EEG-based dynamic affective trajectory prediction.
PeakFlow first learns a coarse affective flow through EEG temporal tokenization and masked temporal modeling, and then applies a lightweight residual refiner to perform peak-guided bounded calibration.
The refiner uses trajectory-aware cues and is optimized by a peak-centered calibration objective that combines global trajectory consistency, peak-zone emphasis, peak-probability localization, terminal suppression, and residual regularization.
This design allows PeakFlow to preserve the global affective trend while correcting peak misalignment, peak-value deviation, and false-terminal peak predictions.
Leave-one-subject-out experiments on SEED-VII show that PeakFlow improves both global trajectory fitting and peak-centered temporal reliability compared with strong dynamic modeling baselines.
Auxiliary evaluation on FIRMED further suggests its potential for sparse peak-centered ordinal intensity analysis.
These results highlight the importance of peak-aware modeling for temporally faithful EEG-based dynamic emotion prediction. Code is available at \url{https://github.com/jukebox333/PeakFlow}.

\end{abstract}

\section{Introduction}

Electroencephalography (EEG)-based emotion recognition has become an important research topic in affective computing and brain--computer interfaces. Compared with external behavioral cues such as facial expression, speech, and gesture, EEG provides a non-invasive measurement of neural activity and is less dependent on overt emotional expression. Existing EEG emotion recognition studies have made substantial progress by designing discriminative spectral features, such as differential entropy (DE), and by developing deep spatial--temporal models for subject-dependent and cross-subject affective decoding~\cite{picard1997affective,koelstra2012deap,duan2013differential,zheng2015investigating,lawhern2018eegnet,song2020dgcnn,zhong2022rgnn}. However, most of these studies formulate affective decoding as a static category prediction problem, where an EEG trial or temporal segment is assigned to a discrete emotion label, such as valence, arousal, or a basic emotion category.

Although static category prediction is useful for recognizing the dominant affective state, it provides only a coarse description of emotional experience.
During continuous stimulation, emotional responses are often not fully captured by a single static label.
Instead, they may gradually accumulate, fluctuate with stimulus content, reach a moment of maximum intensity, and then decay or recover.

This motivates EEG-based dynamic emotion trajectory prediction, where the goal is to estimate a continuous affective intensity curve from sequential EEG observations.
Compared with conventional static classification, this formulation better reflects the temporal nature of emotional responses and enables fine-grained analysis of affective dynamics.

Recent datasets and temporal modeling methods have begun to support this direction.
For example, SEED-VII provides dense temporally resolved affective intensity annotations, making it possible to evaluate whether an EEG model can recover the full evolution of emotional intensity over a trial~\cite{jiang2024seedvii}.
On the methodological side, EEGDancer introduces a dynamic EEG modeling framework that combines vector-quantized temporal representation learning with masked temporal modeling, providing a strong baseline for continuous EEG-based affective trajectory prediction~\cite{zhou2026eegdancer}.
More broadly, temporal representation learning methods, including recurrent networks, temporal convolutional networks, Transformers, vector-quantized representation learning, and masked modeling, provide useful tools for capturing long-range dependencies and learning structured latent representations from EEG sequences~\cite{vaswani2017attention,oord2017vqvae,devlin2019bert,he2022mae}.
While these advances make continuous affective trajectory prediction feasible, existing dynamic EEG models are still mainly designed to recover a globally plausible intensity curve.
Their training objectives and evaluation protocols usually emphasize point-wise fitting quality over the entire trial, such as reducing MSE or improving correlation between predicted and ground-truth trajectories.

However, global trajectory fitting does not necessarily guarantee the correctness of local temporal structures that are crucial for interpreting dynamic emotional experience.
Among these structures, the affective peak is especially important.
It denotes the time point at which emotional intensity reaches its maximum within a trial and corresponds to the strongest emotional response during continuous stimulation.
Accurately identifying such peak moments is important not only for understanding the temporal organization of emotional experience, but also for practical affective computing applications, such as multimedia content analysis, affect-aware human--computer interaction, personalized recommendation, adaptive user-state monitoring, and closed-loop neurofeedback systems.
For example, in multimedia and interactive scenarios, the most emotionally salient moment often determines user engagement, perceived experience, and subsequent behavioral responses.
Psychological studies on retrospective evaluations of affective episodes have also shown that peak and ending moments play important roles in how people evaluate emotional experiences~\cite{fredrickson1993duration,redelmeier1996patients}.
Nevertheless, most point-wise objectives do not explicitly constrain the timing or magnitude of the affective peak.
As a result, a model may capture the coarse affective trend and achieve reasonable global fitting performance while still mislocalizing the most salient affective moment.

\begin{figure}[!t]
\centering
\includegraphics[width=0.98\linewidth]{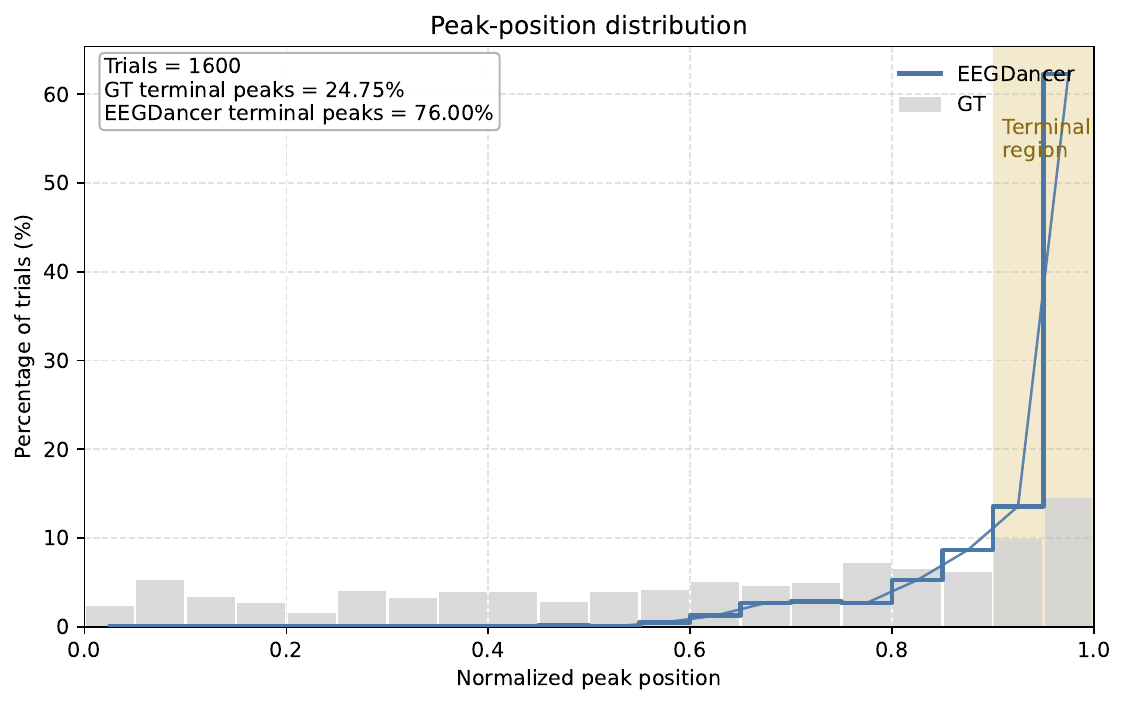}
\caption{
Evidence of terminal-peak bias in SEED-VII.
For each trial, the affective peak is defined as the temporal window with the maximum intensity.
The ground-truth peak positions are broadly distributed across the trial timeline, whereas EEGDancer assigns a much larger proportion of predicted peaks to the terminal region.
The shaded area denotes the terminal region, corresponding to the last 10\% of valid temporal windows. Note that this terminal-peak proportion differs from FTR, which counts only cases where the predicted peak is terminal while the ground-truth peak is non-terminal.
}
\label{fig:peak_position_distribution}
\end{figure}

In our empirical analysis of masked temporal trajectory prediction, we observe a recurring structural failure mode in which the predicted maximum intensity is frequently shifted toward the trial ending, even when the ground-truth affective peak occurs earlier.
We refer to this phenomenon as \textbf{terminal-peak bias}.
As shown in Fig.~\ref{fig:peak_position_distribution}, EEGDancer exhibits a strong terminal concentration of predicted peaks: only 24.75\% of ground-truth peaks occur in the terminal region, whereas 76.00\% of predicted peaks are assigned to this region.

To address this problem, we propose \textbf{PeakFlow}, a peak-guided coarse-to-refined framework for EEG-based dynamic affective trajectory prediction.
PeakFlow decouples global affective-flow modeling from peak-centered structural refinement.
It first learns compact EEG temporal-state tokens through feature-level vector quantization and then trains a masked dynamic Transformer to estimate a globally coherent coarse affective trajectory, where the learned tokens are used as auxiliary code targets.
Given the coarse trajectory, a lightweight residual refiner performs peak-guided bounded calibration rather than replacing the full prediction.
By exploiting trajectory-aware cues, including coarse intensity, normalized temporal position, distance to the terminal region, and local temporal variation, the refiner estimates bounded residual corrections under peak-centered temporal constraints.
This design enables PeakFlow to preserve the global affective trend while correcting peak misalignment, peak-value deviation, local temporal distortion, and false-terminal peak bias.

To train the refinement module, we introduce a peak-centered calibration objective beyond conventional point-wise trajectory fitting.
The objective jointly considers global trajectory consistency, peak-zone emphasis, peak-probability localization, terminal suppression, and residual regularization.
In addition, we argue that conventional global regression metrics are insufficient for evaluating EEG-based dynamic affective trajectory prediction.
Therefore, we evaluate models using both global metrics, including MSE, MAE, Pearson correlation coefficient (PCC), and \(R^2\), and peak-centered metrics, including normalized peak-time error, peak-value error, and false-terminal peak rate.

We conduct leave-one-subject-out (LOSO) experiments on SEED-VII, a dynamic EEG emotion dataset with dense affective intensity annotations.
The results show that PeakFlow improves both global trajectory prediction and peak-centered temporal reliability compared with strong dynamic modeling baselines.
In particular, PeakFlow substantially improves peak localization and peak-value estimation while suppressing false-terminal peak predictions, demonstrating its effectiveness for temporally faithful EEG-based dynamic emotion prediction.
In addition, we perform an auxiliary evaluation on FIRMED to examine whether peak-centered predictions preserve ordinal intensity consistency under sparse event-level annotations.

The main contributions of this work are summarized as follows:

\begin{itemize}

    \item We identify \textbf{terminal-peak bias}, a structural failure mode in masked temporal trajectory prediction where the predicted affective peak is frequently shifted toward the trial ending, revealing the limitation of global point-wise regression metrics.

    \item We propose \textbf{PeakFlow}, a stage-wise coarse-to-refined framework that integrates EEG temporal tokenization, masked affective-flow modeling, and peak-guided bounded residual calibration for dynamic EEG affective trajectory prediction.

    \item We introduce peak-centered evaluation metrics, including normalized peak-time error, peak-value error, and false-terminal peak rate, to assess temporal reliability beyond conventional global regression metrics.

    \item Extensive LOSO experiments on SEED-VII demonstrate that PeakFlow improves both global trajectory prediction and peak-centered temporal reliability, achieving better peak localization, peak-value estimation, and terminal-bias suppression. Auxiliary evaluation on FIRMED further suggests its potential for sparse peak-centered ordinal intensity analysis.

\end{itemize}

\section{Related Work}

\subsection{EEG-Based Emotion Recognition}

EEG-based emotion recognition has been widely studied in affective computing and brain--computer interface research.
Early studies mainly relied on hand-crafted EEG features, such as DE, power spectral density (PSD), hemispheric asymmetry, and functional connectivity, followed by conventional classifiers including support vector machines, \(k\)-nearest neighbors, random forests, and shallow neural networks~\cite{wang2011eeg,duan2013differential,zheng2015investigating}.
These studies demonstrated that EEG signals contain discriminative neural patterns related to affective states, but their performance is often limited by manually designed features and insufficient modeling of spatial--temporal dependencies.

With the development of deep learning, EEG emotion recognition has shifted toward representation learning.
Convolutional neural networks and compact architectures such as EEGNet have been used to extract temporal and spectral EEG patterns~\cite{lawhern2018eegnet}.
Recurrent neural networks, temporal convolutional networks, graph neural networks, and transformer-based models further improve the modeling of sequential dependencies, inter-channel relationships, and long-range temporal interactions~\cite{song2018dgcnn,zhong2020rgnn,vaswani2017attention}.
Another important direction is cross-subject generalization, where domain adaptation, adversarial learning, distribution alignment, and subject-invariant representation learning are used to reduce inter-subject distribution shifts~\cite{ganin2016dann,long2015mmd,sun2016coral}.

Despite these advances, most existing EEG emotion recognition methods still formulate affective decoding as a trial-level or segment-level classification task.
While effective for recognizing dominant affective states, such a formulation does not explicitly characterize how emotional intensity accumulates, changes, peaks, and recovers over time.
This limitation motivates a shift from static category recognition to dynamic affective trajectory modeling.

\subsection{Dynamic EEG Affective Trajectory Prediction}
\label{subsec:eeg_dynamic_emotion_prediction}

Affective responses during continuous multimedia stimulation often exhibit temporal evolution, including gradual accumulation, stimulus-dependent fluctuation, and changes in emotional intensity.
This view is consistent with studies on emotion dynamics and emotional intensity profiles, which emphasize that emotional experiences contain rich temporal variation~\cite{verduyn2009intensity,kuppens2017emotion}.
Therefore, dynamic EEG affective trajectory prediction aims to estimate a continuous emotion intensity curve from sequential EEG observations, providing a more fine-grained description of affective responses than static emotion classification.

Recent EEG emotion datasets have begun to support temporally fine-grained affective modeling.
For example, SEED-VII provides dense continuous affective intensity trajectories at a fixed temporal resolution, enabling trial-level dynamic emotion prediction~\cite{jiang2024seedvii}.
MGEED reflects a similar trend by considering temporally organized emotion elicitation and EEG responses~\cite{wang2023mgeed}.
Different from dense trajectory annotations, FIRMED provides temporally localized affective annotations through an immediate-recall paradigm, where participants report emotionally salient moments and their corresponding intensity levels after stimulus viewing~\cite{tang2025coarse}.
These datasets indicate a clear shift from static emotion recognition toward temporally resolved EEG affective modeling.

From the methodological perspective, recurrent neural networks, temporal convolutional networks, and transformer-based models have been introduced to capture sequential dependencies across EEG windows.
Recent studies further attempt to explicitly model emotional dynamics.
For example, trend-based methods formulate emotion changes as increasing, decreasing, or stable states between adjacent windows and introduce auxiliary trend prediction~\cite{zhang2026time}.
However, trend prediction usually describes only the local direction of emotional change and does not directly recover a continuous affective intensity trajectory.
More recently, EEGDancer combines discrete latent tokenization and masked temporal modeling for EEG-based continuous emotion prediction, providing a strong framework for modeling global affective flow from EEG sequences~\cite{zhou2026eegdancer}.

\subsection{Peak-Centered Affective Trajectory Modeling}
\label{subsec:peak_centered_affective_trajectory}

Beyond global dynamic emotion prediction, peak-centered affective trajectory modeling focuses on whether the most emotionally salient moment can be accurately localized and calibrated.
During continuous multimedia stimulation, emotional intensity may accumulate, fluctuate with stimulus content, reach a local or global peak, and then decrease or recover.
This view is consistent with studies on emotion dynamics and emotion intensity profiles, which suggest that emotional experiences exhibit temporally structured variations rather than remaining constant over time~\cite{kuppens2017emotion,verduyn2009emotion,verduyn2012determinants}.
Moreover, psychological studies on duration neglect, remembered utility, and the peak-end rule show that retrospective evaluations of affective episodes are strongly influenced by salient temporal moments, especially peak intensity and ending states~\cite{varey1992experiences,fredrickson1993duration,kahneman1993more,redelmeier1996patients,ariely2000gestalt,alaybek2022meta}.
Recent annotation paradigms such as FIRMED further highlight the importance of emotionally salient moments through peak-centered immediate-recall annotations~\cite{tang2025coarse}.

However, most existing dynamic EEG affective modeling methods mainly optimize and evaluate trajectory prediction using global regression metrics, such as MSE, MAE, PCC, and \(R^2\).
Although these metrics measure overall point-wise fitting quality, they do not explicitly evaluate whether the predicted trajectory correctly captures the peak time and peak intensity.
As a result, a model may achieve competitive global performance while still failing to recover the most emotionally salient moment.

In addition, temporal models optimized for coarse or smooth trajectory reconstruction, including recurrent networks, temporal convolutional networks, transformer-based models, and masked temporal modeling frameworks~\cite{bai2018tcn,vaswani2017attention,zhou2026eegdancer}, may still produce over-smoothed trajectories or biased peak locations, including terminal-biased peak predictions.
These limitations can lead to unreliable peak-centered interpretation and suggest that conventional global metrics are insufficient for evaluating the temporal structure of dynamic affective trajectories.

\section{Method}
\label{sec:method}

\subsection{Problem Formulation}
\label{subsec:problem_formulation}

We formulate EEG-based dynamic emotion recognition as a trial-wise affective intensity trajectory prediction problem. For the \(i\)-th EEG trial, the recording is segmented into \(T_i\) valid temporal windows. Each window is represented by a \(D\)-dimensional EEG feature vector, resulting in a sequence
\begin{equation}
    \mathbf{X}_i =
    [\mathbf{x}_{i,1},\mathbf{x}_{i,2},\ldots,\mathbf{x}_{i,T_i}]
    \in \mathbb{R}^{T_i\times D}.
\end{equation}
The corresponding normalized affective intensity trajectory and valid-position mask are denoted by
\begin{equation}
    \mathbf{y}_i =
    [y_{i,1},y_{i,2},\ldots,y_{i,T_i}],
    \qquad
    \mathbf{m}_i\in\{0,1\}^{T_i},
\end{equation}
where \(y_{i,t}\in[0,1]\) and \(m_{i,t}=1\) indicates that the \(t\)-th temporal window is valid. The objective is to predict a continuous affective trajectory
\begin{equation}
    \hat{\mathbf{y}}_i =
    [\hat{y}_{i,1},\hat{y}_{i,2},\ldots,\hat{y}_{i,T_i}]
    \in [0,1]^{T_i}.
\end{equation}
To characterize peak-centered temporal structure, we define the ground-truth and predicted affective peaks as
\begin{equation}
    p_i =
    \operatorname*{arg\,max}_{t:m_{i,t}=1} y_{i,t},
    \qquad
    \hat{p}_i =
    \operatorname*{arg\,max}_{t:m_{i,t}=1} \hat{y}_{i,t}.
\end{equation}
PeakFlow aims to preserve the global affective trend while improving peak timing, peak intensity, post-peak dynamics, and false-terminal peak suppression.

\subsection{Overview of Stage-wise PeakFlow}
\label{subsec:stagewise_overview}

\begin{figure*}[!t]
    \centering
    \includegraphics[width=\textwidth]{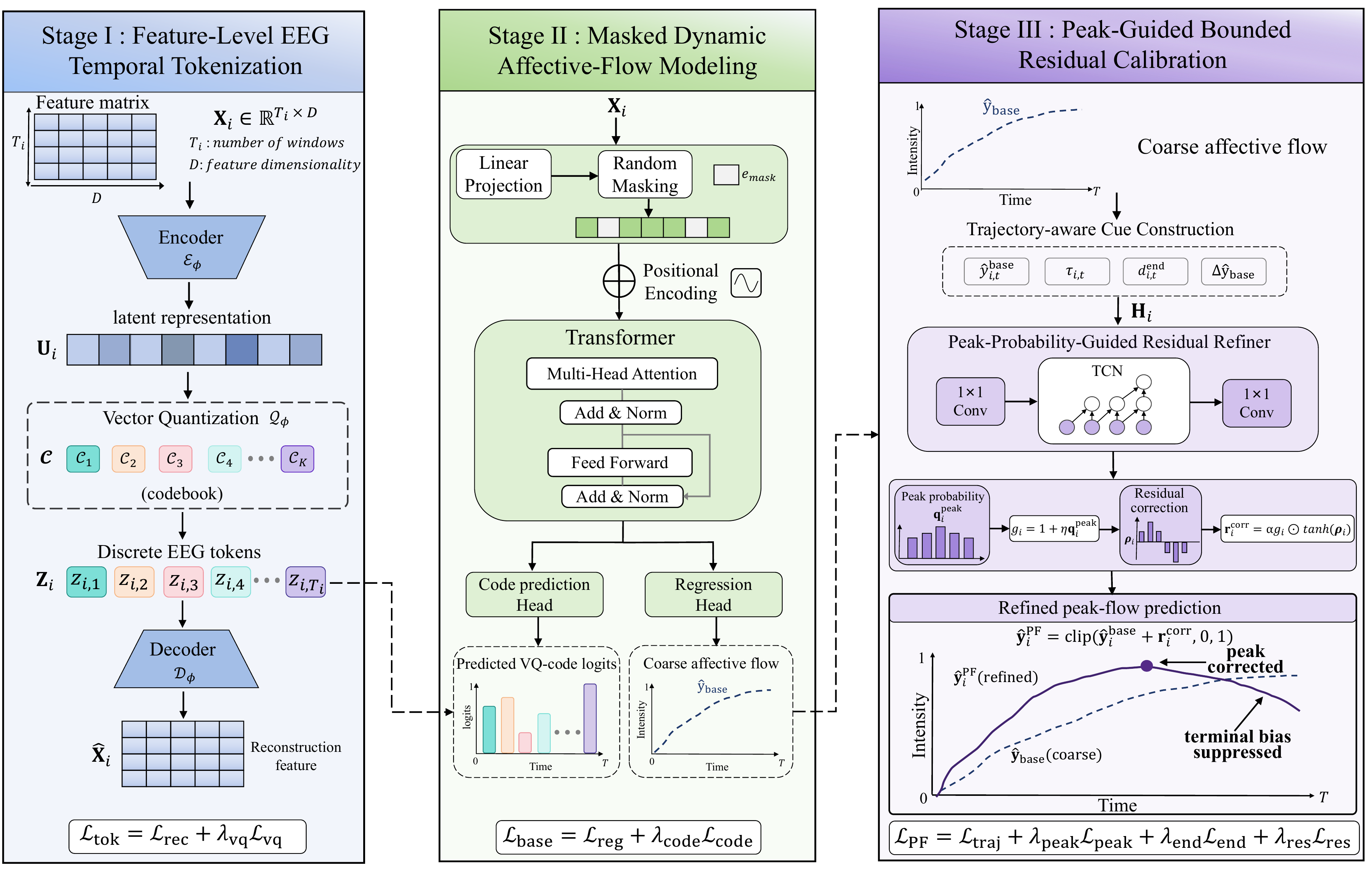}
    \caption{Stage-wise framework of PeakFlow. Stage I learns a feature-level EEG tokenizer with a learnable codebook. Stage II freezes the tokenizer and trains a masked dynamic Transformer for coarse affective-flow prediction. Stage III freezes the previous stages and trains a peak-guided bounded residual calibration module to refine peak timing, peak value, post-peak dynamics, and terminal-peak behavior.}
    \label{fig:peakflow_framework}
\end{figure*}

PeakFlow follows a stage-wise coarse-to-refined design for EEG-based dynamic affective trajectory prediction, as shown in Fig.~\ref{fig:peakflow_framework} and Algorithm~\ref{alg:peakflow}. The framework consists of three sequential stages, each addressing a different level of the prediction problem.

\textbf{Stage I} learns a feature-level EEG temporal tokenizer. It maps continuous EEG window features into discrete temporal-state indices through a learnable codebook. The goal of this stage is not to predict emotion intensity, but to obtain compact EEG temporal states that can provide auxiliary supervision for later dynamic modeling. \textbf{Stage II} learns a coarse affective flow. With the tokenizer frozen, a masked dynamic Transformer predicts the global intensity trajectory from EEG features, while using the discrete temporal-state indices from Stage I as auxiliary code targets. This stage captures the overall temporal evolution of affective intensity, but may still produce over-smoothed peaks or terminal-peak bias. \textbf{Stage III} performs peak-guided bounded residual calibration. With the previous stages frozen, the refiner uses trajectory-aware cues derived from the coarse flow to estimate a soft peak-probability sequence and a bounded residual correction. This stage focuses on correcting peak timing, peak value, and false terminal peaks while preserving the global trend predicted in Stage II.

Overall, PeakFlow separates EEG temporal-state learning, global affective-flow modeling, and local peak-centered refinement into three trainable stages, improving both training stability and interpretability.

\begin{algorithm}[t]
\caption{Stage-wise training and inference of PeakFlow}
\label{alg:peakflow}
\small
\begin{algorithmic}[1]
\Require Training trials $\{(\mathbf{X}_i,\mathbf{y}_i,\mathbf{m}_i)\}_{i=1}^{N}$; peak-zone radius $R$.
\Ensure Final prediction $\hat{\mathbf{y}}^{\mathrm{PF}}_i$.

\Statex \textbf{Stage I: EEG temporal tokenization}
\For{each mini-batch}
    \State Encode EEG features and quantize them with the codebook:
    $\mathbf{U}_i=\mathcal{E}_{\phi}(\mathbf{X}_i)$,
    $(\mathbf{Z}_i,\mathbf{Q}_i)=\mathcal{Q}_{\phi}(\mathbf{U}_i,\mathcal{C})$.
    \State Reconstruct EEG features:
    $\hat{\mathbf{X}}_i=\mathcal{D}_{\phi}(\mathbf{Q}_i)$.
    \State Update the tokenizer by minimizing
    $\mathcal{L}_{\mathrm{tok}}=\mathcal{L}_{\mathrm{rec}}+\lambda_{\mathrm{vq}}\mathcal{L}_{\mathrm{vq}}$.
\EndFor
\State Freeze the Stage-I tokenizer.

\Statex \textbf{Stage II: Masked dynamic affective-flow modeling}
\For{each mini-batch}
    \State Use $\mathbf{X}_i$ as input and the frozen $\mathbf{Z}_i$ as auxiliary code targets.
    \State Apply feature projection, random masking, positional encoding, and Transformer modeling to obtain $\mathbf{H}^{\mathrm{tr}}_i$.
    \State Predict VQ-code logits $\boldsymbol{\ell}_{i,t}$ and coarse affective flow $\hat{\mathbf{y}}^{\mathrm{base}}_i$.
    \State Update the Stage-II predictor by minimizing
    $\mathcal{L}_{\mathrm{base}}=\mathcal{L}_{\mathrm{reg}}+\lambda_{\mathrm{code}}\mathcal{L}_{\mathrm{code}}$.
\EndFor
\State Freeze the Stage-II predictor.

\Statex \textbf{Stage III: Peak-guided bounded residual calibration}
\For{each mini-batch}
    \State Obtain coarse flow $\hat{\mathbf{y}}^{\mathrm{base}}_i$ from the frozen Stage-II predictor.
    \State Construct trajectory-aware cues $\mathbf{H}_i$ and peak-zone labels
    $q^{\mathrm{zone}}_{i,t}=\mathbf{1}[|t-p_i|\leq R]$.
    \State Use the TCN refiner to predict peak logits $\mathbf{a}_i$ and raw residual scores $\boldsymbol{\rho}_i$.
    \State Compute
    $\mathbf{q}^{\mathrm{peak}}_i=\sigma(\mathbf{a}_i)$,
    $\mathbf{g}_i=1+\eta\mathbf{q}^{\mathrm{peak}}_i$,
    $\mathbf{r}^{\mathrm{corr}}_i=\alpha\mathbf{g}_i\odot\tanh(\boldsymbol{\rho}_i)$.
    \State Obtain
    $\hat{\mathbf{y}}^{\mathrm{PF}}_i=
    \operatorname{clip}(\hat{\mathbf{y}}^{\mathrm{base}}_i+\mathbf{r}^{\mathrm{corr}}_i,0,1)$.
    \State Update only the Stage-III refiner by minimizing $\mathcal{L}_{\mathrm{PF}}$.
\EndFor

\Statex \textbf{Inference}
\State Given a test trial $\mathbf{X}_i$, predict $\hat{\mathbf{y}}^{\mathrm{base}}_i$, construct $\mathbf{H}_i$, estimate $\mathbf{q}^{\mathrm{peak}}_i$ and $\mathbf{r}^{\mathrm{corr}}_i$, and return $\hat{\mathbf{y}}^{\mathrm{PF}}_i$.
\end{algorithmic}
\end{algorithm}

\subsection{Stage I: Feature-Level EEG Temporal Tokenization}
\label{subsec:stage1_tokenization}

The first stage learns compact discrete EEG temporal states from continuous EEG features. Given \(\mathbf{X}_i\), a temporal encoder maps each EEG window into a latent representation:
\begin{equation}
    \mathbf{U}_i =
    \mathcal{E}_{\phi}(\mathbf{X}_i)
    =
    [\mathbf{u}_{i,1},\ldots,\mathbf{u}_{i,T_i}]
    \in \mathbb{R}^{T_i\times d_z}.
\end{equation}
In the implementation, \(\mathcal{E}_{\phi}\) is a two-layer MLP:
\begin{equation}
    \mathbf{u}_{i,t}
    =
    \mathbf{W}^{(2)}_e
    \operatorname{Dropout}
    \left(
    \operatorname{GELU}
    \left(
    \mathbf{W}^{(1)}_e\mathbf{x}_{i,t}
    +
    \mathbf{b}^{(1)}_e
    \right)
    \right)
    +
    \mathbf{b}^{(2)}_e .
\end{equation}
For SEED-VII DE features, the default input dimension is \(D=310\), the hidden dimension is \(128\), and the latent dimension is \(d_z=64\).

The codebook is maintained inside the vector quantizer. It contains \(K\) learnable prototype embeddings:
\begin{equation}
    \mathcal{C}
    =
    \{\mathbf{c}_1,\mathbf{c}_2,\ldots,\mathbf{c}_{K}\},
    \qquad
    \mathbf{c}_{k}\in\mathbb{R}^{d_z}.
\end{equation}
In the default setting, \(K=64\). For each latent EEG window embedding \(\mathbf{u}_{i,t}\), the nearest codeword is selected by squared Euclidean distance:
\begin{equation}
    z_{i,t}
    =
    \operatorname*{arg\,min}_{k}
    \left\|
    \mathbf{u}_{i,t}
    -
    \mathbf{c}_{k}
    \right\|_2^2 .
\end{equation}
The quantized embedding is
\begin{equation}
    \mathbf{q}_{i,t}
    =
    \mathbf{c}_{z_{i,t}},
\end{equation}
and the discrete EEG temporal-state sequence is
\begin{equation}
    \mathbf{Z}_i
    =
    [z_{i,1},z_{i,2},\ldots,z_{i,T_i}].
\end{equation}
During back-propagation, the straight-through estimator is adopted:
\begin{equation}
    \tilde{\mathbf{q}}_{i,t}
    =
    \mathbf{u}_{i,t}
    +
    \operatorname{sg}
    \left[
    \mathbf{q}_{i,t}
    -
    \mathbf{u}_{i,t}
    \right],
\end{equation}
where \(\operatorname{sg}[\cdot]\) denotes the stop-gradient operation.

A decoder reconstructs the EEG feature vector from the quantized embedding:
\begin{equation}
    \hat{\mathbf{x}}_{i,t}
    =
    \mathcal{D}_{\phi}(\tilde{\mathbf{q}}_{i,t}).
\end{equation}
The tokenizer is trained by
\begin{equation}
    \mathcal{L}_{\mathrm{tok}}
    =
    \mathcal{L}_{\mathrm{rec}}
    +
    \lambda_{\mathrm{vq}}
    \mathcal{L}_{\mathrm{vq}},
\end{equation}
where
\begin{equation}
    \mathcal{L}_{\mathrm{rec}}
    =
    \frac{
    \sum_{i,t}m_{i,t}
    \left\|
    \mathbf{x}_{i,t}
    -
    \hat{\mathbf{x}}_{i,t}
    \right\|_2^2
    }{
    \sum_{i,t}m_{i,t}
    },
\end{equation}
and
\begin{equation}
    \mathcal{L}_{\mathrm{vq}}
    =
    \left\|
    \operatorname{sg}[\mathbf{U}_i]
    -
    \mathbf{Q}_i
    \right\|_2^2
    +
    \beta
    \left\|
    \mathbf{U}_i
    -
    \operatorname{sg}[\mathbf{Q}_i]
    \right\|_2^2 .
\end{equation}
After Stage I training, \(\mathcal{E}_{\phi}\), \(\mathcal{Q}_{\phi}\), the codebook \(\mathcal{C}\), and the decoder are frozen. The learned token sequence \(\mathbf{Z}_i\) is used as auxiliary discrete-state supervision in Stage II.

\subsection{Stage II: Masked Dynamic Affective-Flow Modeling}
\label{subsec:stage2_masked_dynamic_modeling}

The second stage learns a globally coherent coarse affective flow. Importantly, the masked dynamic Transformer uses the continuous EEG feature sequence \(\mathbf{X}_i\) as input, while the discrete EEG tokens \(\mathbf{Z}_i\) learned in Stage I are used as auxiliary code targets. This avoids relying solely on discrete tokens and preserves continuous EEG information for affective regression.

Given \(\mathbf{X}_i\), each EEG window is first projected into a hidden space:
\begin{equation}
    \mathbf{s}_{i,t}
    =
    \mathbf{W}_{p}\mathbf{x}_{i,t}
    +
    \mathbf{b}_{p},
    \qquad
    \mathbf{s}_{i,t}\in\mathbb{R}^{d_h}.
\end{equation}
A subset of valid temporal positions is randomly masked during training. Let \(b_{i,t}\in\{0,1\}\) denote the random masking indicator. The masked hidden representation is
\begin{equation}
    \tilde{\mathbf{s}}_{i,t}
    =
    \begin{cases}
    \mathbf{e}_{\mathrm{mask}}, & b_{i,t}=1,\\
    \mathbf{s}_{i,t}, & b_{i,t}=0,
    \end{cases}
\end{equation}
where \(\mathbf{e}_{\mathrm{mask}}\in\mathbb{R}^{d_h}\) is a learnable mask token. Positional encoding is then added:
\begin{equation}
    \mathbf{o}_{i,t}
    =
    \tilde{\mathbf{s}}_{i,t}
    +
    \mathbf{p}_{t}.
\end{equation}
The resulting sequence is processed by a Transformer encoder:
\begin{equation}
    \mathbf{H}^{\mathrm{tr}}_i
    =
    \operatorname{Transformer}_{\theta}
    \left(
    \mathbf{O}_i,
    \mathbf{m}_i
    \right),
\end{equation}
where invalid padded positions are ignored by the key-padding mask. The implemented Transformer uses \(3\) encoder layers, \(4\) attention heads, hidden dimension \(128\), feed-forward dimension \(512\), GELU activation, dropout \(0.1\), batch-first computation, and pre-normalization.

The Transformer output is fed into two heads. The code prediction head estimates the VQ code index:
\begin{equation}
    \boldsymbol{\ell}_{i,t}
    =
    \mathbf{W}_{c}\mathbf{h}^{\mathrm{tr}}_{i,t}
    +
    \mathbf{b}_{c},
    \qquad
    \boldsymbol{\ell}_{i,t}\in\mathbb{R}^{K}.
\end{equation}
The regression head predicts the coarse affective intensity:
\begin{equation}
    \hat{y}^{\mathrm{base}}_{i,t}
    =
    \sigma
    \left(
    \mathbf{w}^{\top}_{r}
    \operatorname{GELU}
    \left(
    \mathbf{W}_{r}
    \operatorname{LN}
    \left(
    \mathbf{h}^{\mathrm{tr}}_{i,t}
    \right)
    +
    \mathbf{b}_{r}
    \right)
    +
    b_{r}
    \right).
\end{equation}
Thus, the coarse affective flow is
\begin{equation}
    \hat{\mathbf{y}}^{\mathrm{base}}_i
    =
    [
    \hat{y}^{\mathrm{base}}_{i,1},
    \ldots,
    \hat{y}^{\mathrm{base}}_{i,T_i}
    ].
\end{equation}

The Stage II objective combines continuous trajectory regression and discrete state prediction:
\begin{equation}
    \mathcal{L}_{\mathrm{base}}
    =
    \mathcal{L}_{\mathrm{reg}}
    +
    \lambda_{\mathrm{code}}
    \mathcal{L}_{\mathrm{code}},
\end{equation}
where
\begin{equation}
    \mathcal{L}_{\mathrm{reg}}
    =
    \frac{
    \sum_{i,t}m_{i,t}
    \left|
    \hat{y}^{\mathrm{base}}_{i,t}
    -
    y_{i,t}
    \right|
    }{
    \sum_{i,t}m_{i,t}
    },
\end{equation}
and
\begin{equation}
    \mathcal{L}_{\mathrm{code}}
    =
    \frac{
    \sum_{i,t}m_{i,t}
    \operatorname{CE}
    \left(
    \boldsymbol{\ell}_{i,t},
    z_{i,t}
    \right)
    }{
    \sum_{i,t}m_{i,t}
    }.
\end{equation}
During Stage II training, the Stage I tokenizer is frozen and only the masked dynamic Transformer and its prediction heads are updated. After training, the coarse model is frozen. The output \(\hat{\mathbf{y}}^{\mathrm{base}}_i\) provides a global affective trend for Stage III.

\subsection{Stage III: Peak-Guided Bounded Residual Calibration}
\label{subsec:stage3_peak_guided_refinement}

After Stage II, the masked dynamic Transformer provides a globally coherent but coarse affective flow
\(\hat{\mathbf{y}}^{\mathrm{base}}_i\). Although this coarse prediction captures the overall temporal trend, it may still suffer from peak misalignment, peak-value underestimation, over-smoothed dynamics, and false terminal peaks. Therefore, Stage III performs peak-guided bounded residual calibration on top of the coarse trajectory. During this stage, the Stage I tokenizer and the Stage II coarse predictor are frozen, and only the residual calibration module is trained.

\subsubsection{Trajectory-Aware Cue Construction}
\label{subsubsec:trajectory_aware_cue}

The Stage III refiner does not directly take raw EEG features as input. Instead, it operates on trajectory-level cues derived from the coarse affective flow. For the \(i\)-th trial, the coarse predictions are first grouped according to subject and video identifiers and sorted by temporal indices. For each valid temporal window, we construct a compact trajectory-aware cue vector:
\begin{equation}
\mathbf{h}_{i,t}
=
\left[
    \hat{y}^{\mathrm{base}}_{i,t},
    \tau_{i,t},
    d^{\mathrm{end}}_{i,t},
    \Delta \hat{y}^{\mathrm{base}}_{i,t}
\right],
\label{eq:stage3_cue}
\end{equation}
where
\begin{equation}
    \tau_{i,t}
    =
    \frac{t-1}{T_i-1},
    \qquad
    d^{\mathrm{end}}_{i,t}
    =
    1-\tau_{i,t},
\end{equation}
and
\begin{equation}
    \Delta \hat{y}^{\mathrm{base}}_{i,t}
    =
    \hat{y}^{\mathrm{base}}_{i,t}
    -
    \hat{y}^{\mathrm{base}}_{i,t-1}.
\end{equation}
Here, \(\hat{y}^{\mathrm{base}}_{i,t}\) provides the coarse intensity anchor, \(\tau_{i,t}\) encodes the normalized temporal position, \(d^{\mathrm{end}}_{i,t}\) serves as a terminal-aware positional cue, and \(\Delta \hat{y}^{\mathrm{base}}_{i,t}\) represents the local rising or falling tendency of the coarse flow. The cue sequence is denoted as
\begin{equation}
    \mathbf{H}_i
    =
    [\mathbf{h}_{i,1},\ldots,\mathbf{h}_{i,T_i}]
    \in
    \mathbb{R}^{T_i\times C}.
\end{equation}
Compared with using only the coarse intensity value, the explicit positional and terminal-distance cues help the refiner distinguish genuine affective peaks from false terminal responses.

For peak-probability supervision, we define a binary peak-zone label according to the ground-truth peak position:
\begin{equation}
    q^{\mathrm{zone}}_{i,t}
    =
    \mathbf{1}
    \left[
    |t-p_i|\leq R
    \right],
\label{eq:peak_zone_label}
\end{equation}
where \(R\) is the peak-zone radius. This label is used only during training.

\subsubsection{Peak-Probability-Guided Residual Refiner}
\label{subsubsec:peak_guided_refiner}

The cue sequence \(\mathbf{H}_i\) is first projected into a hidden space by a \(1\times1\) convolution:
\begin{equation}
    \mathbf{S}_i
    =
    \operatorname{Conv}_{1\times1}
    \left(
    \mathbf{H}_i
    \right).
\end{equation}
Then, a lightweight temporal convolutional network models local and mid-range trajectory patterns:
\begin{equation}
    \mathbf{V}^{(0)}_i
    =
    \mathbf{S}_i,
    \qquad
    \mathbf{V}^{(\ell)}_i
    =
    \operatorname{TCNBlock}_{\ell}
    \left(
    \mathbf{V}^{(\ell-1)}_i
    \right),
    \quad
    \ell=1,\ldots,L .
\end{equation}
Each TCN block consists of one-dimensional temporal convolutions, batch normalization, GELU activation, dropout, and residual addition. The dilation rate is set as \(d_{\ell}=2^{\ell-1}\), which enlarges the temporal receptive field while keeping the calibration module lightweight.

The final hidden representation is shared by two prediction heads. The residual head predicts a raw correction score, while the peak head predicts a peak logit:
\begin{equation}
\begin{aligned}
    \boldsymbol{\rho}_i
    &=
    \operatorname{Conv}^{\mathrm{res}}_{1\times1}
    \left(
    \mathbf{V}^{(L)}_i
    \right),\\
    \mathbf{a}_i
    &=
    \operatorname{Conv}^{\mathrm{peak}}_{1\times1}
    \left(
    \mathbf{V}^{(L)}_i
    \right).
\end{aligned}
\end{equation}
The soft peak-probability sequence is obtained by
\begin{equation}
    \mathbf{q}^{\mathrm{peak}}_i
    =
    \sigma(\mathbf{a}_i).
\label{eq:qpeak}
\end{equation}
Instead of directly adding \(\mathbf{q}^{\mathrm{peak}}_i\) to the predicted intensity, PeakFlow uses it as a soft gate to modulate the residual correction:
\begin{equation}
    \mathbf{g}_i
    =
    1+\eta\mathbf{q}^{\mathrm{peak}}_i,
\end{equation}
\begin{equation}
    \mathbf{r}^{\mathrm{corr}}_i
    =
    \alpha
    \mathbf{g}_i
    \odot
    \tanh
    \left(
    \boldsymbol{\rho}_i
    \right),
\label{eq:bounded_residual_qpeak}
\end{equation}
where \(\alpha\) controls the maximum residual scale and \(\eta\) controls the strength of peak-guided modulation. The final PeakFlow prediction is
\begin{equation}
    \hat{\mathbf{y}}^{\mathrm{PF}}_i
    =
    \operatorname{clip}
    \left(
    \hat{\mathbf{y}}^{\mathrm{base}}_i
    +
    \mathbf{r}^{\mathrm{corr}}_i,
    0,1
    \right).
\label{eq:stage3_final_prediction}
\end{equation}
In this design, \(\mathbf{q}^{\mathrm{peak}}_i\) indicates where peak-centered calibration should be emphasized, while \(\boldsymbol{\rho}_i\) determines how the trajectory should be locally adjusted. The bounded residual formulation prevents the refiner from destroying the global trend learned in Stage II.

\subsubsection{Peak-Centered Calibration Objective}
\label{subsubsec:stage3_objective}

Stage III is optimized with a peak-centered calibration objective:
\begin{equation}
\mathcal{L}_{\mathrm{PF}}
=
\mathcal{L}_{\mathrm{traj}}
+
\lambda_{\mathrm{peak}}\mathcal{L}_{\mathrm{peak}}
+
\lambda_{\mathrm{end}}\mathcal{L}_{\mathrm{end}}
+
\lambda_{\mathrm{res}}\mathcal{L}_{\mathrm{res}} .
\label{eq:stage3_loss}
\end{equation}
This objective jointly considers trajectory fitting, peak localization, terminal-peak suppression, and bounded residual regularization.

The trajectory term preserves global fitting and first-order temporal consistency:
\begin{equation}
\mathcal{L}_{\mathrm{traj}}
=
\mathcal{L}_{\mathrm{fit}}
+
\omega_{\Delta}\mathcal{L}_{\Delta},
\end{equation}
where
\begin{equation}
\mathcal{L}_{\mathrm{fit}}
=
\frac{
\sum_{i,t}m_{i,t}
\left(\hat{y}^{\mathrm{PF}}_{i,t}-y_{i,t}\right)^2
}{
\sum_{i,t}m_{i,t}
}.
\end{equation}
Here, \(\mathcal{L}_{\Delta}\) matches the first-order temporal differences between
\(\hat{\mathbf{y}}^{\mathrm{PF}}_i\) and \(\mathbf{y}_i\) over adjacent valid windows.

The peak term consists of peak-zone fitting and peak-probability localization:
\begin{equation}
\mathcal{L}_{\mathrm{peak}}
=
\mathcal{L}_{\mathrm{pz}}
+
\omega_{\mathrm{prob}}\mathcal{L}_{\mathrm{prob}} .
\end{equation}
To avoid over-emphasizing non-peak regions, the peak-zone fitting term applies a larger weight around the annotated peak:
\begin{equation}
\mathcal{L}_{\mathrm{pz}}
=
\frac{
\sum_{i,t}m_{i,t}w_{i,t}e_{i,t}^{2}
}{
\sum_{i,t}m_{i,t}
},
\label{eq:stage3_peak_zone_loss}
\end{equation}
where
\begin{equation}
e_{i,t}
=
\hat{y}^{\mathrm{PF}}_{i,t}
-
y_{i,t},
\qquad
w_{i,t}
=
1+
(\omega_{\mathrm{pz}}-1)q^{\mathrm{zone}}_{i,t}.
\end{equation}
The peak-probability supervision is defined as
\begin{equation}
\mathcal{L}_{\mathrm{prob}}
=
\frac{
\sum_{i,t}m_{i,t}
\operatorname{BCE}
\left(
q^{\mathrm{peak}}_{i,t},
q^{\mathrm{zone}}_{i,t}
\right)
}{
\sum_{i,t}m_{i,t}
}.
\end{equation}

The terminal term penalizes over-estimated responses in the last valid windows:
\begin{equation}
L_{\mathrm{end}}
=
\frac{1}{N}
\sum_i
\frac{1}{|\mathcal{T}^{\mathrm{term}}_i|}
\sum_{t\in \mathcal{T}^{\mathrm{term}}_i}
\left(
\left[
\hat{y}^{\mathrm{PF}}_{i,t}
-
y_{i,t}
\right]_{+}
\right)^2 .
\end{equation}

where \([a]_+ = \max(a,0)\), and \(\mathcal{T}^{\mathrm{term}}_i\) denotes the terminal region defined as the last \(r_{\mathrm{term}}\) proportion of valid temporal windows.
In all experiments, we set \(r_{\mathrm{term}}=0.10\).

Finally, the residual regularization limits unnecessary correction:
\begin{equation}
\mathcal{L}_{\mathrm{res}}
=
\frac{
\sum_{i,t}m_{i,t}
\left(r^{\mathrm{corr}}_{i,t}\right)^2
}{
\sum_{i,t}m_{i,t}
}.
\end{equation}
Overall, the objective encourages accurate trajectory reconstruction while explicitly emphasizing peak localization, terminal-bias suppression, and stable residual calibration.

\subsection{Training and Inference}
\label{subsec:training_inference}

PeakFlow is optimized following the three-stage procedure described above. In Stage I, the feature-level tokenizer is pretrained using \(\mathcal{L}_{\mathrm{tok}}\). In Stage II, the tokenizer is frozen and the masked dynamic Transformer is trained using \(\mathcal{L}_{\mathrm{base}}\). In Stage III, both the tokenizer and coarse predictor are frozen, and the peak-guided residual calibration module is trained using \(\mathcal{L}_{\mathrm{PF}}\). The implementation uses AdamW optimization, gradient clipping, and early stopping according to validation performance.

During inference, only the EEG feature sequence and valid mask are required. The trained coarse predictor first produces \(\hat{\mathbf{y}}^{\mathrm{base}}_i\). PeakFlow then constructs trajectory-level cues, estimates the peak probability, predicts the bounded residual correction, and obtains the final trajectory according to Eq.~\eqref{eq:stage3_final_prediction}. The final prediction is evaluated using both global regression metrics and peak-centered metrics, including normalized peak-time error, peak-value error, and false-terminal peak rate.

\section{Experiments}
\label{sec:experiments}

\subsection{Datasets and Experimental Settings}
\label{subsec:datasets_settings}

\textbf{SEED-VII\cite{jiang2024seedvii}.}
We conduct the main experiments on SEED-VII, a dynamic EEG emotion dataset with temporally resolved affective intensity annotations. Different from conventional EEG emotion recognition datasets that provide only trial-level emotion labels, SEED-VII contains continuous affective intensity annotations at a fixed temporal resolution, making it suitable for evaluating dynamic affective trajectory prediction. Following the dataset protocol, EEG signals are segmented into temporal windows, and differential entropy (DE) features are extracted from multiple frequency bands. For the \(i\)-th trial, the EEG feature sequence is denoted as \(\mathbf{X}_i\in\mathbb{R}^{T_i\times D}\), and the corresponding normalized affective intensity trajectory is denoted as \(\mathbf{y}_i\in[0,1]^{T_i}\). Since different trials may have different valid lengths, all sequences are padded to the maximum sequence length and a binary mask \(\mathbf{m}_i\in\{0,1\}^{T_i}\) is used to indicate valid temporal positions.

\textbf{FIRMED\cite{tang2025coarse}.}
To further examine the applicability of PeakFlow under sparse peak-centered affective supervision, we additionally use FIRMED for auxiliary cross-dataset analysis. Different from SEED-VII, which provides dense continuous intensity trajectories, FIRMED contains sparse event-level affective annotations with event timestamps, emotion categories, and ordered intensity levels. Therefore, FIRMED is not used as a direct dense-trajectory benchmark. Instead, it is used to evaluate whether the proposed peak-aware formulation can preserve ordered affective intensity semantics under sparse peak-centered annotations.

\textbf{Experimental protocol.}
We adopt the Leave-One-Subject-Out (LOSO) protocol to evaluate subject-independent generalization. In each fold, all trials from one subject are used as the test set, while trials from the remaining subjects are used for training and validation. This subject-independent protocol avoids window-level leakage and ensures that the test subject is unseen during training. 

\textbf{Implementation details.}
PeakFlow is trained in a coarse-to-refined manner following the three-stage procedure described in Section~\ref{sec:method}. First, the feature-level EEG temporal tokenizer is pretrained to learn compact discrete temporal-state tokens from EEG feature sequences. Second, the masked dynamic Transformer is trained to predict a globally coherent coarse affective flow, with the tokenizer frozen and the learned token indices used as auxiliary code targets. Third, the peak-guided residual refiner is trained on top of the frozen coarse predictor to calibrate peak timing, peak value, local temporal dynamics, and false-terminal responses. 

During training, invalid padded positions are excluded by the binary mask. The model is optimized using AdamW with early stopping based on validation performance. Hyperparameters, including the codebook size, Transformer configuration, masking ratio, residual scale, and loss weights, are selected on the validation set and kept fixed for all test folds. 

\subsection{Evaluation Metrics}
\label{subsec:evaluation_metrics}

We evaluate PeakFlow from three complementary perspectives: global trajectory fitting, peak-centered reliability, and terminal-peak bias.

\textbf{Global trajectory metrics.}
We report standard regression metrics, including mean squared error (MSE), mean absolute error (MAE), Pearson correlation coefficient (PCC), and the coefficient of determination \(R^2\). These metrics measure the overall fitting quality between the predicted affective trajectory \(\hat{\mathbf{y}}_i\) and the ground-truth trajectory \(\mathbf{y}_i\).

\textbf{Peak-centered metrics.}
Since dynamic affective prediction should preserve the most intense emotional moment, we further evaluate peak localization and peak intensity. For the \(i\)-th trial, the ground-truth and predicted peak positions are defined as
\begin{equation}
    p_i = \operatorname*{arg\,max}_{t:m_{i,t}=1} y_{i,t},
    \qquad
    \hat{p}_i = \operatorname*{arg\,max}_{t:m_{i,t}=1} \hat{y}_{i,t}.
\end{equation}
We report normalized peak-time error and peak-value error:
\begin{equation}
    E_{\mathrm{time}}
    =
    \frac{1}{N}
    \sum_i
    \frac{|\hat{p}_i-p_i|}{T_i-1},
    \qquad
    E_{\mathrm{value}}
    =
    \frac{1}{N}
    \sum_i
    \left|
    \hat{y}_{i,\hat{p}_i}
    -
    y_{i,p_i}
    \right|.
\end{equation}

\textbf{Terminal-peak bias metrics.}
To quantify whether a model incorrectly shifts the predicted peak toward the end of a trial, we define the terminal region as the last \(r_{\mathrm{term}}\) proportion of valid temporal windows, where \(r_{\mathrm{term}}=0.10\) by default.
For the \(i\)-th trial with \(T_i\) valid temporal windows, the number of terminal windows is defined as
\begin{equation}
K_i^{\mathrm{term}}
=
\left\lceil r_{\mathrm{term}} T_i \right\rceil .
\end{equation}
The terminal region is then defined as
\begin{equation}
\mathcal{T}^{\mathrm{term}}_i
=
\left\{
t \mid T_i - K_i^{\mathrm{term}} + 1 \le t \le T_i
\right\}.
\end{equation}
A false-terminal peak occurs when the predicted peak falls in this region while the ground-truth peak does not:
\begin{equation}
I^{\mathrm{false}}_i
=
\mathbf{1}
\left[
\hat{p}_i \in \mathcal{T}^{\mathrm{term}}_i
\wedge
p_i \notin \mathcal{T}^{\mathrm{term}}_i
\right].
\end{equation}
The false-terminal peak rate is then computed as
\begin{equation}
\mathrm{FTR}
=
\frac{1}{N}
\sum_{i=1}^{N}
I^{\mathrm{false}}_i .
\end{equation}
A lower FTR indicates fewer artificial terminal peaks and more reliable peak localization.

\textbf{Ordinal intensity-level metrics.}
For FIRMED and auxiliary intensity-level analysis, continuous predictions are discretized into ordered intensity levels. We report macro-F1, ordinal mean absolute error (Ord. MAE), and quadratic weighted kappa (QWK). These metrics are used only for auxiliary validation under sparse event-level supervision and are not directly compared with SEED-VII dense trajectory metrics.

\subsection{Overall Performance on SEED-VII}
\label{subsec:overall_performance_seedvii}

Table~\ref{tab:seedvii_overall_results} reports the main results on SEED-VII under the LOSO protocol.
Since SEED-VII provides dense continuous affective intensity trajectories, we evaluate both global regression quality and peak-centered temporal structure.
We include conventional window-wise regression baselines, deep temporal sequence models, the EEGDancer baseline, and the final PeakFlow.

\begin{table*}[!t]
\centering
\caption{
Main results on SEED-VII for dense continuous affective trajectory prediction under the LOSO protocol.
Window-wise methods predict each EEG window independently, while temporal modeling methods predict trial-level affective trajectories.
FTR denotes false-terminal peak rate.
}
\label{tab:seedvii_overall_results}
\small
\renewcommand{\arraystretch}{1.18}
\setlength{\tabcolsep}{4.4pt}
\begin{tabular}{p{0.23\textwidth}ccccccc}
\toprule
\multirow{2}{*}{Method}
& \multicolumn{4}{c}{Global Metrics}
& \multicolumn{3}{c}{Peak-Centered Metrics} \\
\cmidrule(lr){2-5} \cmidrule(lr){6-8}
& MSE $\downarrow$ & MAE $\downarrow$ & PCC $\uparrow$ & $R^2$ $\uparrow$
& Peak-Time $\downarrow$ & Peak-Value $\downarrow$ & FTR $\downarrow$ \\
\midrule
\textit{Window-wise regression} & & & & & & & \\
Ridge Regression
& 0.0964 & 0.2608 & 0.3746 & 0.0185
& 0.4632 & 0.2697 & 38.75\% \\
SVR
& 0.0918 & 0.2521 & 0.4103 & 0.0468
& 0.4375 & 0.2584 & 41.31\% \\
MLP
& 0.0872 & 0.2436 & 0.4569 & 0.0715
& 0.4086 & 0.2462 & 45.94\% \\
\midrule
\textit{Temporal modeling} & & & & & & & \\
GRU
& 0.0829 & 0.2358 & 0.4937 & 0.0924
& 0.3817 & 0.2351 & 51.63\% \\
TCN
& 0.0806 & 0.2319 & 0.5148 & 0.1036
& 0.3664 & 0.2293 & 54.81\% \\
Transformer
& 0.0789 & 0.2287 & 0.5316 & 0.1119
& 0.3528 & 0.2258 & 57.44\% \\
EEGDancer
& 0.0758 & 0.2234 & 0.5523 & 0.1172
& 0.3409 & 0.2219 & 62.56\% \\
\midrule
PeakFlow
& \textbf{0.0733} & \textbf{0.2149} & \textbf{0.5754} & \textbf{0.1411}
& \textbf{0.2658} & \textbf{0.2005} & \textbf{5.42\%} \\
\bottomrule
\end{tabular}
\vspace{-1.5mm}
\end{table*}

Ridge Regression, SVR, and MLP predict affective intensity at each EEG window independently.
GRU, TCN, and Transformer Encoder are sequence-to-sequence regression models that predict the full affective trajectory.
EEGDancer introduces discrete temporal tokenization and masked temporal modeling, and is therefore used as a strong dynamic trajectory prediction baseline.
The final PeakFlow further introduces peak-probability-guided bounded residual calibration and terminal suppression.

\begin{figure*}[!t]
\centering
\includegraphics[width=0.93\linewidth,
height=0.31\textheight,
keepaspectratio]{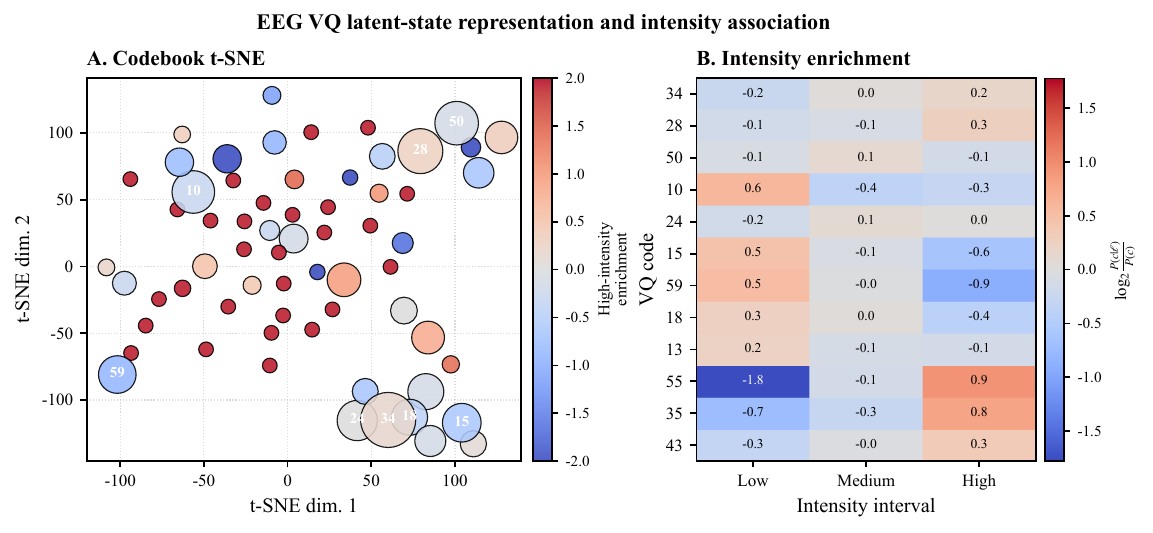}
\caption{
Visualization of the learned EEG VQ latent states on SEED-VII.
(A) The VQ codebook embeddings are projected into a two-dimensional space using t-SNE.
Marker size denotes the usage frequency of each code, and color indicates its high-intensity enrichment tendency.
(B) Intensity-conditioned enrichment of representative VQ codes across low-, medium-, and high-intensity intervals, measured by \(\log_2(P(c|\ell)/P(c))\).
The results suggest that the learned EEG latent states exhibit non-uniform usage patterns and intensity-dependent associations.
}
\label{fig:vq_latent_states}
\vspace{-1.5mm}
\end{figure*}

\paragraph{Analysis of learned EEG latent states}
Since EEGDancer learns discrete EEG temporal states through the VQ-VAE tokenizer, and PeakFlow further refines the affective flow built upon this dynamic representation, we visualize the learned VQ latent states in Fig.~\ref{fig:vq_latent_states}.
The t-SNE projection of the codebook embeddings shows that the learned codes form diverse latent prototypes with non-uniform usage frequencies.
Moreover, the intensity-conditioned enrichment analysis indicates that several frequently activated codes exhibit different preferences across low-, medium-, and high-intensity intervals.
This suggests that the VQ tokenizer captures structured EEG latent states related to affective intensity evolution, providing a meaningful discrete representation for subsequent masked temporal modeling.

\paragraph{Overall comparison}
Based on this structured dynamic representation, PeakFlow further improves peak-centered trajectory prediction.
Compared with EEGDancer, PeakFlow reduces MSE from 0.0758 to 0.0733 and MAE from 0.2234 to 0.2149, while increasing PCC from 0.5523 to 0.5754.
More importantly, PeakFlow reduces normalized peak-time MAE from 0.3409 to 0.2658, peak-value MAE from 0.2219 to 0.2005, and false-terminal peak rate from 62.56\% to 5.42\%.
These results indicate that PeakFlow improves not only point-wise trajectory fitting but also the reliability of affective peak localization and terminal-bias suppression.

\begin{figure*}[!t]
    \centering
    \includegraphics[width=0.93\textwidth]{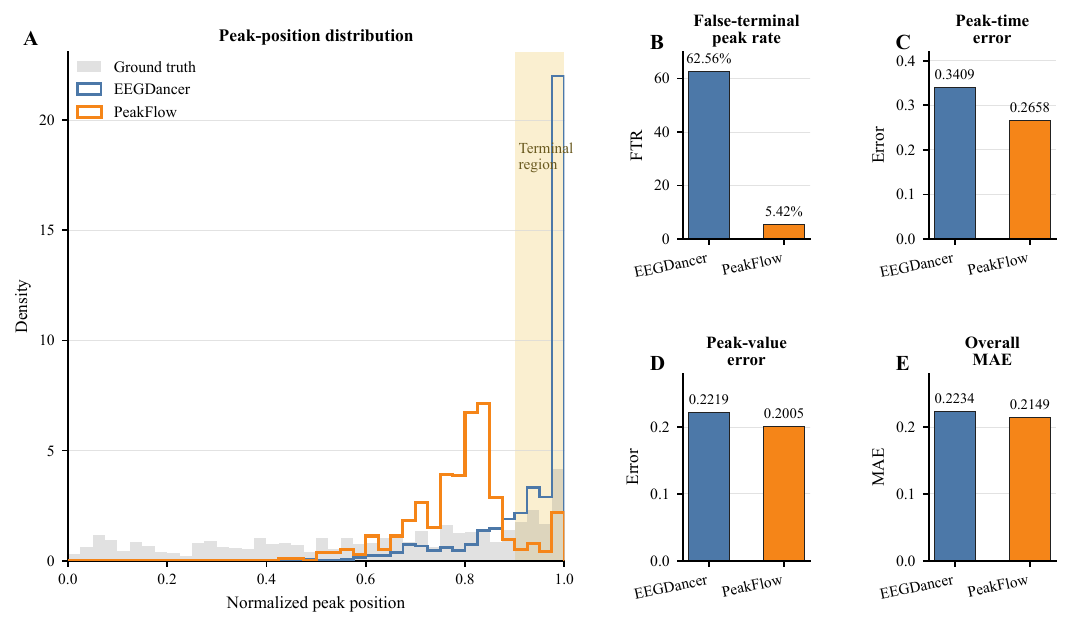}
    \caption{
    Peak-centered analysis and terminal-peak bias correction on SEED-VII.
    (A) Distribution of normalized peak positions.
    (B) False-terminal peak rate.
    (C) Normalized peak-time error.
    (D) Peak-value error.
    (E) Overall MAE.
    The terminal region is consistently defined as the last \(r_{\mathrm{term}}=10\%\) of valid temporal windows.
    }
    \label{fig:peak_centered_analysis}
\end{figure*}

\paragraph{Peak-centered analysis}
In addition to the overall regression metrics, we further examine whether PeakFlow improves peak-centered temporal prediction and alleviates the terminal-peak bias.
As shown in Fig.~\ref{fig:peak_centered_analysis}, EEGDancer tends to shift the predicted peak toward the terminal region of a trial, resulting in a substantially higher false-terminal peak rate.
In contrast, PeakFlow produces a peak-position distribution that is more consistent with the ground truth and greatly reduces terminal false peaks.
These results indicate that the proposed peak-aware refinement not only improves global trajectory fitting, but also better preserves the temporal structure around affective peaks.

\begin{figure*}[!t]
\centering
\includegraphics[
width=0.98\textwidth,
height=0.50\textheight,
keepaspectratio
]{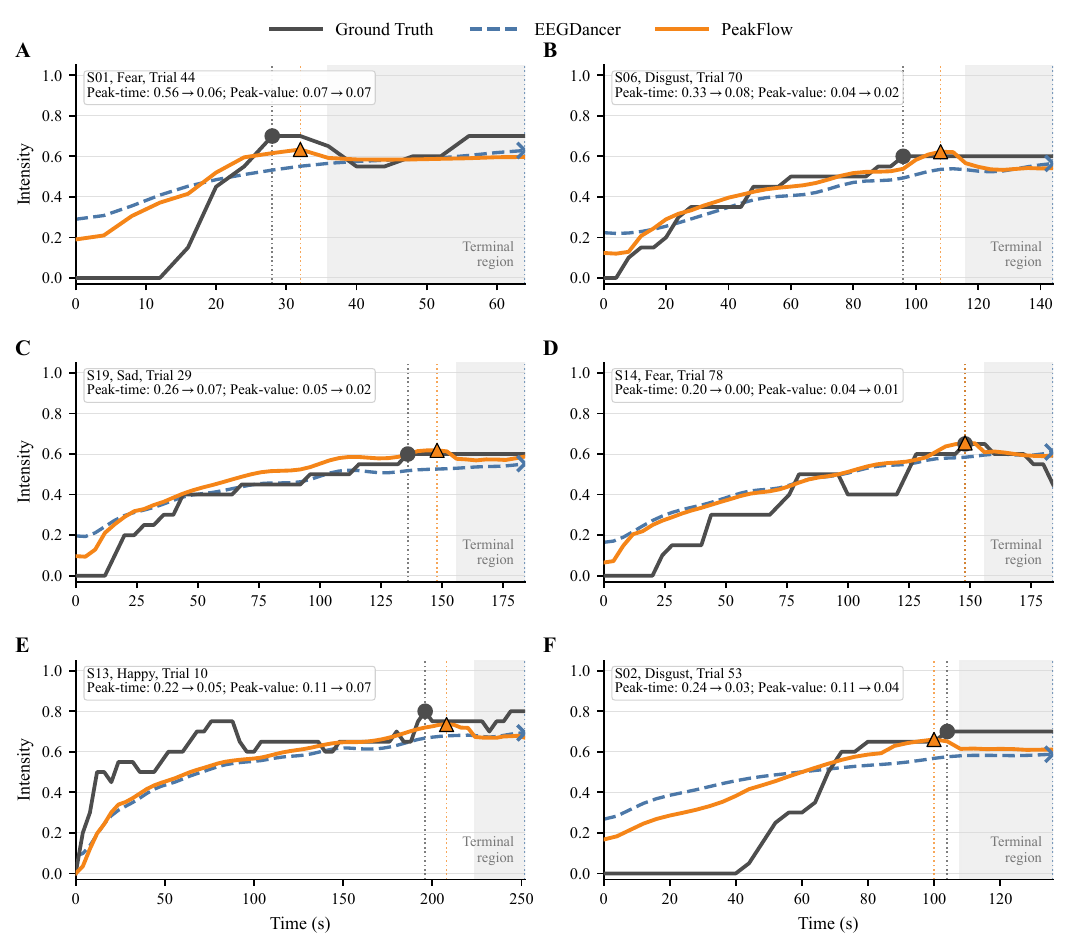}
\caption{
Qualitative examples of peak-aware trajectory refinement on SEED-VII.
Each subplot shows the ground-truth affective intensity trajectory, the EEGDancer prediction, and the PeakFlow prediction for a representative trial.
The shaded area denotes the terminal region.
Peak-time and peak-value errors are reported as EEGDancer~$\rightarrow$~PeakFlow.
}
\label{fig:peakflow_cases}
\vspace{-1.5mm}
\end{figure*}

\paragraph{Qualitative trajectory analysis}
To provide a more intuitive understanding of the peak-aware refinement behavior, we further visualize representative trial-level trajectories in Fig.~\ref{fig:peakflow_cases}.
Across different subjects and emotion categories, EEGDancer can capture the coarse affective trend, but it often shifts the predicted peak toward the terminal region or produces inaccurate peak timing and terminal-biased peak locations.
In contrast, PeakFlow better aligns the predicted peak with the ground-truth trajectory while preserving the overall affective flow.
These qualitative cases indicate that the proposed refiner mainly improves peak timing correction, terminal-bias suppression, and local trajectory refinement.

\begin{figure}[!b]
\centering
\includegraphics[
width=1\linewidth,
height=0.4\textheight,
keepaspectratio
]{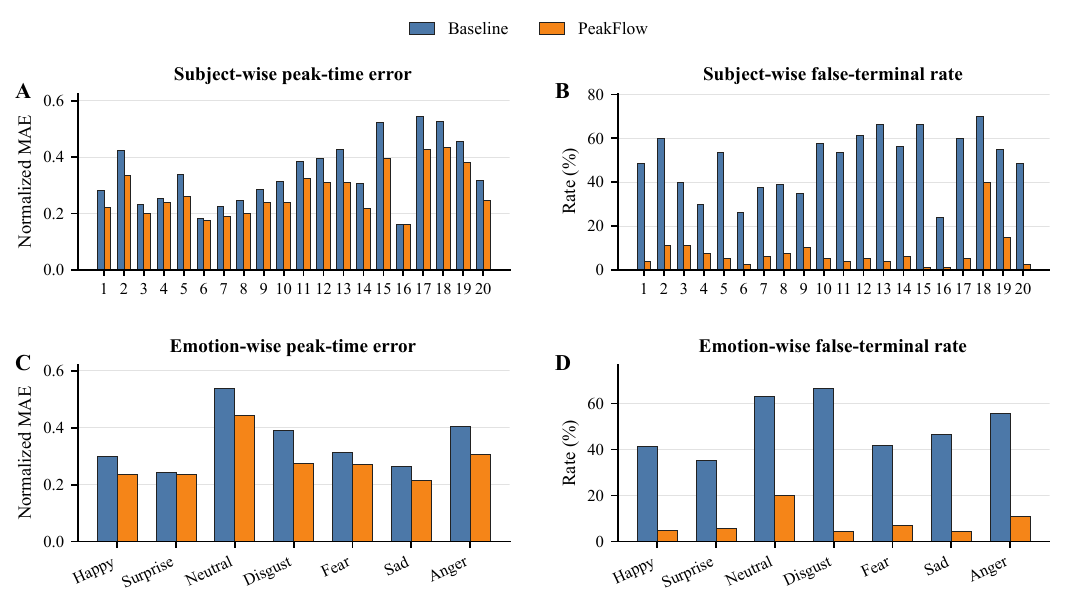}
\caption{
Subject-wise and emotion-wise consistency analysis on SEED-VII.
(A) Subject-wise normalized peak-time error.
(B) Subject-wise false-terminal peak rate.
(C) Emotion-wise normalized peak-time error.
(D) Emotion-wise false-terminal peak rate.
}
\label{fig:subject_emotion_consistency}
\vspace{-1.5mm}
\end{figure}

\paragraph{Subject- and emotion-wise consistency}
We further examine whether the peak-centered improvements are consistent across subjects and emotion categories.
As shown in Fig.~\ref{fig:subject_emotion_consistency}, PeakFlow reduces normalized peak-time error and false-terminal peak rate for most subjects.
Similar improvements can also be observed across emotion categories, suggesting that the proposed refinement is not limited to a specific subject or affective class.
These results demonstrate that PeakFlow provides more stable peak-centered temporal modeling under cross-subject EEG variability.

\subsection{Ablation and Hyperparameter Sensitivity}
\label{subsec:ablation_sensitivity}

To further examine the contribution of each component in PeakFlow, we conduct an ablation study by removing one design at a time. The evaluated variants include removing the peak-centered objective, removing the bounded residual design, removing the terminal penalty, and removing temporal cues from the refiner input. The results are reported in Table~\ref{tab:ablation_peakflow}.

\begin{table*}[!t]
\centering
\caption{
Ablation study of PeakFlow on SEED-VII under the LOSO protocol.
FTR denotes the false-terminal peak rate.
}
\label{tab:ablation_peakflow}
\small
\renewcommand{\arraystretch}{1.18}
\setlength{\tabcolsep}{4.4pt}
\begin{tabular}{p{0.27\textwidth}ccccccc}
\toprule
\multirow{2}{*}{Variant}
& \multicolumn{4}{c}{Global Metrics}
& \multicolumn{3}{c}{Peak-Centered Metrics} \\
\cmidrule(lr){2-5} \cmidrule(lr){6-8}
& MSE $\downarrow$ & MAE $\downarrow$ & PCC $\uparrow$ & $R^2$ $\uparrow$
& Peak-Time $\downarrow$ & Peak-Value $\downarrow$ & FTR $\downarrow$ \\
\midrule
Full PeakFlow
& \textbf{0.0733}
& \textbf{0.2149}
& \textbf{0.5754}
& \textbf{0.1411}
& \textbf{0.2658}
& \textbf{0.2005}
& \textbf{5.42\%} \\

w/o peak-centered objective
& 0.0738
& 0.2158
& 0.5709
& 0.1354
& 0.3152
& 0.2091
& 41.44\% \\

w/o bounded residual
& 0.0737
& 0.2160
& 0.5711
& 0.1360
& 0.2765
& 0.2106
& 8.41\% \\

w/o terminal penalty
& 0.0736
& 0.2157
& 0.5720
& 0.1375
& 0.2814
& 0.2107
& 18.01\% \\

w/o temporal cues
& 0.0741
& 0.2172
& 0.5667
& 0.1322
& 0.2786
& 0.2121
& 15.99\% \\
\bottomrule
\end{tabular}
\vspace{-1.5mm}
\end{table*}

The ablation results show that the peak-centered objective plays a crucial role in correcting the peak structure of the predicted affective flow. Removing this objective substantially increases the peak-time error from 0.2658 to 0.3152 and the false-terminal peak rate from 5.42\% to 41.44\%. It also degrades the peak-value error from 0.2005 to 0.2091 and reduces PCC from 0.5754 to 0.5709. These results indicate that the peak-centered objective is essential for accurate peak localization and terminal-bias correction, supporting our motivation that dynamic affective intensity prediction should explicitly consider peak-related temporal structures rather than only optimizing global regression accuracy.

The terminal penalty also contributes clearly to suppressing spurious terminal peaks. Without this penalty, the false-terminal peak rate increases from 5.42\% to 18.01\%, confirming that terminal-aware regularization helps prevent the model from over-assigning the maximum intensity to the end of a trial. Removing the bounded residual design produces moderate degradation in peak-centered metrics, increasing the peak-time error from 0.2658 to 0.2765 and the peak-value error from 0.2005 to 0.2106, suggesting that the bounded residual correction acts as a stabilizing mechanism that prevents uncontrolled trajectory refinement. Removing temporal cues also degrades the peak-time error, peak-value error, and false-terminal rate, indicating that temporal context is helpful for both peak calibration and terminal-bias suppression. Overall, these results demonstrate that the full PeakFlow model achieves the best performance across both global and peak-centered metrics, and that its improvement mainly comes from enhanced peak localization, bounded residual correction, and explicit suppression of terminal-peak bias.

\begin{figure*}[!t]
    \centering
    \includegraphics[width=0.98\textwidth]{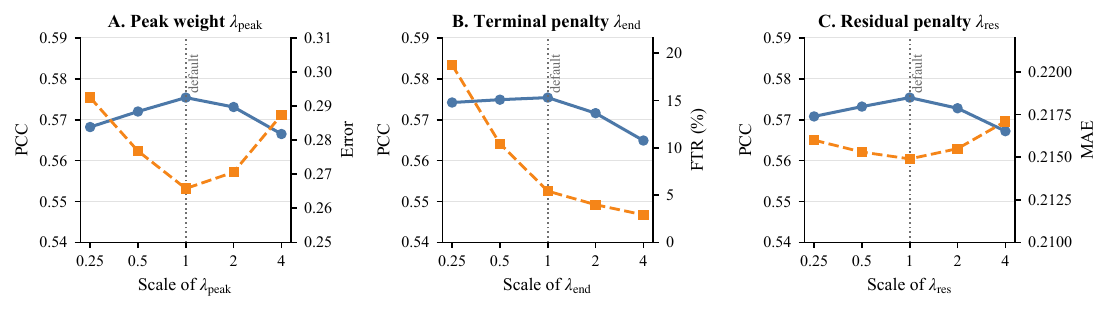}
    \caption{
    Hyperparameter sensitivity analysis of PeakFlow with respect to the Stage-III loss weights.
    We vary the peak weight \(\lambda_{\mathrm{peak}}\), terminal penalty \(\lambda_{\mathrm{end}}\), and residual penalty \(\lambda_{\mathrm{res}}\).
    The blue curve reports PCC, while the orange curve reports the most relevant metric for each weight: peak-time error for \(\lambda_{\mathrm{peak}}\), false-terminal peak rate for \(\lambda_{\mathrm{end}}\), and overall MAE for \(\lambda_{\mathrm{res}}\).
    The vertical dotted line denotes the default setting used in the main experiments.
    }
    \label{fig:hyperparam_sensitivity}
    \vspace{-1.5mm}
\end{figure*}

Fig.~\ref{fig:hyperparam_sensitivity} further evaluates the sensitivity of PeakFlow to three Stage-III loss weights, including the peak weight \(\lambda_{\mathrm{peak}}\), terminal penalty \(\lambda_{\mathrm{end}}\), and residual penalty \(\lambda_{\mathrm{res}}\). Overall, PeakFlow remains stable across a relatively wide range of weight scales. For \(\lambda_{\mathrm{peak}}\), the default setting achieves the best trade-off between global fitting and peak localization, yielding the highest PCC together with the lowest peak-time error. For \(\lambda_{\mathrm{end}}\), increasing the terminal penalty further reduces the false-terminal peak rate, but overly large values lead to a slight decrease in PCC, indicating a trade-off between terminal-bias suppression and global trajectory fitting. For \(\lambda_{\mathrm{res}}\), the default setting also provides the best balance, achieving the highest PCC and the lowest overall MAE, while both smaller and larger values cause mild degradation. Therefore, the default configuration is adopted as a balanced setting in all main experiments. This analysis further suggests that the proposed loss design is not overly sensitive to a narrow hyperparameter choice.

\subsection{Auxiliary Evaluation on FIRMED}
\label{subsec:firmed_auxiliary}

To further examine whether PeakFlow is compatible with sparse peak-centered affective annotations, we conduct an auxiliary evaluation on FIRMED.
Unlike SEED-VII, FIRMED does not provide dense continuous affective trajectories for every temporal window.
Therefore, dense trajectory metrics such as MSE, MAE, PCC, peak-time error, and false-terminal peak rate are not directly comparable in this setting.
Instead, we evaluate ordinal intensity-level consistency under sparse event-level annotations using Macro-F1, ordinal MAE, and quadratic weighted kappa (QWK).

For each annotated FIRMED event, we extract the predicted peak intensity within its event-centered temporal window:
\begin{equation}
\hat{s}_{i,k}
=
\max_{t \in \mathcal{W}_{i,k}}
\hat{y}_{i,t},
\end{equation}
where \(\mathcal{W}_{i,k}\) denotes the temporal window centered at the \(k\)-th annotated affective event of the \(i\)-th trial.
The predicted event-level peak intensity \(\hat{s}_{i,k}\) is then discretized into three ordered intensity levels using the same protocol for EEGDancer and PeakFlow.
Specifically, low, medium, and high intensity levels correspond to the intervals \([0,0.30]\), \((0.30,0.70]\), and \((0.70,1.00]\), respectively.
The sparse FIRMED annotations are mapped to the same ordinal space, where intensity 20 is treated as low, 40 and 60 as medium, and 80 and 100 as high.
This event-centered evaluation avoids treating FIRMED as a dense trajectory benchmark and instead assesses whether the predicted local peak intensity is consistent with the annotated ordinal affective intensity.

\begin{figure}[!t]
\centering
\includegraphics[width=\linewidth]{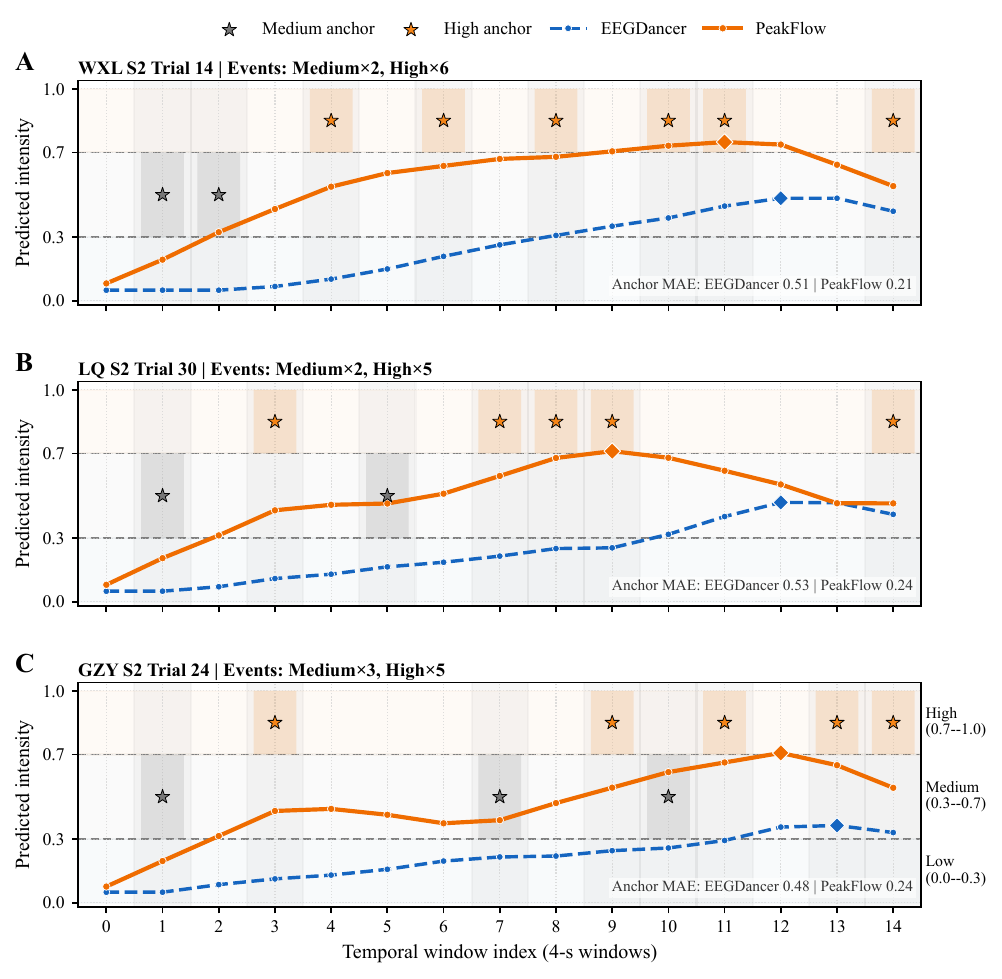}
\caption{
Auxiliary ordinal intensity-level visualization on FIRMED.
The horizontal axis denotes the temporal window index, and the vertical axis denotes the predicted intensity.
Colored anchors indicate sparse event-level ordinal annotations, and dashed horizontal lines mark the boundaries between low, medium, and high intensity intervals.
Compared with EEGDancer, PeakFlow produces trajectories whose event-centered peak intensities are more consistent with sparse ordinal intensity annotations.
}
\label{fig:firmed_trialwise}
\end{figure}

The FIRMED evaluation should be interpreted as a complementary validation rather than a replacement for the SEED-VII dense trajectory benchmark.
SEED-VII evaluates whether a model can recover the full affective intensity trajectory, whereas FIRMED examines whether peak-centered predictions preserve ordered intensity semantics under sparse event-level annotations.
The two evaluations therefore reflect different aspects of dynamic affective modeling: dense trajectory recovery and sparse peak-centered ordinal consistency.

\begin{table}[!t]
\centering
\caption{
Auxiliary ordinal intensity-level evaluation on FIRMED.
Macro-F1, Ord. MAE, and QWK denote macro-averaged F1 score, ordinal mean absolute error, and quadratic weighted kappa, respectively.
}
\label{tab:firmed_ordinal_results}
\small
\renewcommand{\arraystretch}{1.25}
\setlength{\tabcolsep}{5pt}
\begin{tabular}{lccc}
\toprule
Method & Macro-F1 $\uparrow$ & Ord. MAE $\downarrow$ & QWK $\uparrow$ \\
\midrule
EEGDancer & 0.527 & 0.437 & 0.457 \\
PeakFlow & \textbf{0.761} & \textbf{0.231} & \textbf{0.760} \\
\bottomrule
\end{tabular}
\end{table}

As shown in Table~\ref{tab:firmed_ordinal_results}, PeakFlow achieves higher Macro-F1 and QWK and lower ordinal MAE than EEGDancer.
These results suggest that the peak-guided refinement improves ordinal intensity-level consistency under sparse peak-centered annotations.
Together with the SEED-VII dense trajectory results, the FIRMED evaluation further supports the effectiveness of PeakFlow for peak-aware dynamic affective modeling.

\section{Discussion}
\label{sec:discussion}

\subsection{Peak-Aware Refinement Beyond Point-Wise Fitting}
\label{subsec:discussion_refinement}

PeakFlow is effective because it decouples coarse affective-flow modeling from peak-centered structural refinement.
The masked temporal model first estimates a globally coherent affective trajectory, while the peak-aware refiner performs bounded residual calibration around structurally important temporal regions.
This design allows PeakFlow to correct the predicted affective peak without replacing the entire trajectory or disrupting the global temporal trend.

The results support this interpretation.
Compared with EEGDancer, PeakFlow reduces MSE from 0.0758 to 0.0733 and MAE from 0.2234 to 0.2149, while increasing PCC from 0.5523 to 0.5754.
These improvements indicate that the refinement module does not sacrifice global point-wise fitting.
More importantly, PeakFlow substantially improves peak-centered temporal reliability.
It reduces normalized peak-time error from 0.3409 to 0.2658, peak-value error from 0.2219 to 0.2005, and false-terminal peak rate from 62.56\% to 5.42\%.
This suggests that the main advantage of PeakFlow lies not only in reducing average regression error, but also in improving the structural correctness of the predicted affective trajectory.

These findings further show that conventional point-wise metrics are insufficient for evaluating dynamic affective trajectory prediction.
A model may obtain reasonable MSE, MAE, or PCC while still assigning the maximum affective intensity to an incorrect temporal location.
As shown in Fig.~\ref{fig:peak_centered_analysis}, EEGDancer tends to shift predicted peaks toward the terminal region, whereas PeakFlow substantially suppresses this false-terminal behavior.
Therefore, peak-time error, peak-value error, and false-terminal peak rate are necessary complements to conventional global regression metrics for evaluating temporally faithful EEG-based affective trajectory prediction.

\subsection{Consistency Across Subjects and Emotion Categories}
\label{subsec:discussion_consistency}

The improvement of PeakFlow is not limited to a specific subject or emotion category.
Under the LOSO protocol, PeakFlow consistently reduces peak-time error and false-terminal peak rate for most test subjects, indicating that terminal-peak bias is a common failure mode in cross-subject EEG trajectory prediction.
This is important because EEG responses vary substantially across individuals due to physiological differences, neural variability, and subject-specific affective sensitivity.
The subject-wise consistency suggests that the proposed peak-aware refiner can generalize to unseen subjects rather than simply overfitting to particular training subjects.

Emotion-wise results lead to a similar conclusion.
Different emotion categories may exhibit different temporal profiles, such as rapid intensity escalation, gradual accumulation, sustained activation, or relatively flat low-intensity patterns.
Nevertheless, PeakFlow improves peak-centered prediction across emotion categories, suggesting that the proposed refinement mechanism is not merely correcting an emotion-specific artifact.
Instead, it addresses a more general peak-localization problem in dynamic affective modeling.
This observation supports the use of peak-aware trajectory refinement as a general strategy for EEG-based dynamic emotion prediction.

The ablation study further explains the source of these improvements.
Removing the peak-centered objective leads to a clear degradation in peak-time error and false-terminal peak rate, showing that explicitly modeling peak-centered temporal structure is essential.
Removing terminal suppression also increases false-terminal predictions, confirming that terminal bias cannot be fully resolved by global trajectory fitting alone.
In addition, removing bounded residual calibration or temporal cues weakens peak-centered performance, suggesting that stable local correction and trajectory-aware information are both important for reliable affective peak refinement.

\subsection{Complementary FIRMED Evidence and Limitations}
\label{subsec:discussion_limitations}

FIRMED provides complementary evidence rather than a direct replacement for SEED-VII.
SEED-VII contains dense continuous affective intensity trajectories and is therefore suitable for evaluating trajectory-level metrics, including MSE, MAE, PCC, peak-time error, peak-value error, and false-terminal peak rate.
In contrast, FIRMED provides sparse event-centered ordinal intensity annotations rather than dense continuous trajectories.
Therefore, dense trajectory metrics are not directly applicable to FIRMED.
Instead, FIRMED is used for auxiliary ordinal consistency analysis, examining whether peak-centered predictions preserve ordered intensity semantics under sparse event-level supervision.
In this sense, SEED-VII evaluates dense temporal recovery, while FIRMED evaluates sparse peak-centered ordinal consistency.

Several limitations remain.
First, although SEED-VII enables systematic dense trajectory evaluation, more datasets with dense temporal affective annotations are needed to further verify the generality of PeakFlow.
Second, the current framework is designed for offline prediction and uses full-trial temporal context.
Online affective peak anticipation would require causal temporal modeling and uncertainty-aware prediction.
Third, the current study focuses on EEG signals, while affective peaks may also be reflected in facial expression, EDA, ECG, respiration, speech, and other behavioral or physiological modalities.
Extending PeakFlow to multimodal affective trajectory modeling is therefore a promising future direction.
Finally, the current formulation focuses on the strongest affective peak within each trial.
Future work may extend PeakFlow to multi-peak modeling, including onset, apex, offset, and recovery stages of dynamic emotional responses.

\section{Conclusion}
\label{sec:conclusion}

We proposed PeakFlow, a peak-aware masked temporal modeling framework for EEG-based dynamic affective trajectory prediction.
PeakFlow first learns a coarse affective flow through discrete EEG temporal representation learning and masked temporal modeling, and then refines the trajectory using a lightweight peak-guided bounded residual module.
LOSO experiments on SEED-VII demonstrate that PeakFlow improves both global trajectory prediction and peak-centered temporal reliability compared with strong dynamic modeling baselines.
In particular, PeakFlow achieves better affective peak localization, peak-value estimation, and false-terminal peak suppression.
These findings suggest that dynamic EEG emotion recognition should move beyond static labels and global point-wise regression toward peak-aware affective trajectory modeling.

\bibliographystyle{IEEEtran}
\bibliography{references}

\end{document}